\documentclass[11pt]{article} 
\usepackage{graphicx}
\usepackage{hyperref}
\usepackage{amssymb}
\usepackage{authblk}

\date{\today}

\title{Constraints on the neutrino emission from the Galactic Ridge with the ANTARES telescope}

\author[1]{S.~Adri\'an-Mart\'inez}
\author[2]{A.~Albert}
\author[3]{M.~Andr\'e}
\author[4]{M.~Anghinolfi}
\author[5]{G.~Anton}
\author[1]{M.~Ardid}
\author[6]{J.-J.~Aubert}
\author[7]{T.~Avgitas}
\author[7]{B.~Baret}
\author[8]{J.~Barrios-Mart\'{\i}}
% ...

\affil[1]{\scriptsize{Institut d'Investigaci\'o per a la Gesti\'o Integrada de les Zones Costaneres (IGIC) - Universitat Polit\`ecnica de Val\`encia. C/  Paranimf 1 , 46730 Gandia, Spain.}}
\affil[2]{\scriptsize{GRPHE - Universit\'e de Haute Alsace - Institut universitaire de technologie de Colmar, 34 rue du Grillenbreit BP 50568 - 68008 Colmar, France}}
\affil[3]{\scriptsize{Technical University of Catalonia, Laboratory of Applied Bioacoustics, Rambla Exposici\'o,08800 Vilanova i la Geltr\'u,Barcelona, Spain}}
\affil[4]{\scriptsize{INFN - Sezione di Genova, Via Dodecaneso 33, 16146 Genova, Italy}}
\affil[5]{\scriptsize{Friedrich-Alexander-Universit\"at Erlangen-N\"urnberg, Erlangen Centre for Astroparticle Physics, Erwin-Rommel-Str. 1, 91058 Erlangen, Germany}}
\affil[6]{\scriptsize{Aix-Marseille Universit\'e, CNRS/IN2P3, CPPM UMR 7346, 13288 Marseille, France}}
\affil[7]{\scriptsize{APC, Universit\'e Paris Diderot, CNRS/IN2P3, CEA/IRFU, Observatoire de Paris, Sorbonne Paris Cit\'e, 75205 Paris, France}}
\affil[8]{\scriptsize{IFIC - Instituto de F\'isica Corpuscular c/ Catedra\'atico Jos\'e Beltr\'an, 2 E-46980 Paterna, Valencia, Spain}}
% ...

\begin{document} 


\maketitle 

\begin{abstract}

Compelling evidence for the existence of astrophysical neutrinos has been reported by the IceCube collaboration.
Some features of the energy and declination distributions of IceCube events hint at a North/South asymmetry of the neutrino flux. This could be due to the presence of the bulk of our Galaxy in the Southern hemisphere. The ANTARES neutrino telescope, located in the Mediterranean Sea, has been taking data since 2007. It offers the best sensitivity to muon neutrinos produced by galactic cosmic ray interactions in this region of the sky. In this letter a search for an extended neutrino flux from the Galactic Ridge region is presented. Different models of neutrino production by cosmic ray propagation are tested. No excess of events is observed and upper limits for different neutrino flux spectral indices are set. 
This constrains the number of IceCube events possibly originating from the Galactic Ridge. A simple power-law extrapolation of the Fermi-LAT flux to associated IceCube High Energy Starting Events is excluded at 90\% confidence level.

\end{abstract}

\section{Introduction} \label{sec:intro}

Neutrino telescopes search for high-energy neutrinos produced by astrophysical objects. The ANTARES detector \cite{bib:antares} is the largest underwater neutrino telescope. Its effective area, good angular resolution and good exposure to the Southern sky has allowed the detector to produce the best limits on neutrino emission from point-like objects at low declinations \cite{bib:antares_ps}.

The existence of astrophysical neutrinos has been reported by the IceCube collaboration in various analyses \cite{bib:IC2y, bib:IC3y, bib:IC1T,bib:ICmu}. The observed flux is compatible with the hypothesis of isotropy \cite{bib:ICcombi} and equipartition in the three neutrino flavours \cite{bib:ICflavor}. Multiple conjectures have been made for the origin of the observed signal and its features, some of which invoke a possible asymmetry in the Northern vs Southern sky. 
Since the IceCube analysis most-sensitive to the diffuse flux \cite{bib:IC2y} is based on vetoing techniques to detect contained downward-going events \cite{bib:ICveto}, the observed signal is dominantly composed of shower-type events from the Southern Hemisphere, making the analysis relatively insensitive to such an asymmetry. The directional resolution in the High Energy Starting Events (HESE) sample for shower events is limited, and individual sources of the signal have not been identified so far. For the ANTARES location, in the Mediterranean Sea, the Southern sky is accessible via upward-going muon tracks, for which a sub-degree angular resolution is achieved \cite{anta12}. Consequently, an ensemble of individual point sources, or an extended region with an enhanced emission, might appear diffuse to IceCube but resolvable by ANTARES.

The central part of the Milky Way is considered a guaranteed extended source of neutrinos that originate from the decay of short-lived particles produced by the interaction of primary cosmic rays (CRs) with the surrounding matter. Though this flux is usually considered too low to explain the IceCube signal observed in the Southern sky \cite{bib:murase}, recent computations have shown that a possible enhancement of the neutrino flux could be present in the bulk of our Galaxy \cite{bib:antonio, bib:fresh}.

In this letter, a search for neutrinos ($\nu_\mu+\bar{\nu}_\mu$) from the central region of the Galactic Plane is presented, using ANTARES data from 2007 to 2013. The motivation for this search in the context of the IceCube measurement is discussed in \S\,\ref{sec:gal_nu} and the ANTARES detector and the considered data set are introduced in \S\,\ref{sec:detector}. The search method and the optimisation procedures based on Monte Carlo simulations are described in \S\,\ref{sec:data_analysis}. The results of the analysis are presented and discussed in \S\,\ref{sec:results} and \S\,\ref{sec:outlook}.

\section{Neutrinos from our Galaxy and the IceCube signal} \label{sec:gal_nu}

The IceCube data currently provide the sole observation of high-energy cosmic neutrinos. The first measurements \cite{bib:IC2y,bib:IC3y,bib:IC1T} were consistent with an isotropic distribution of the arrival directions and thus an extragalactic origin. The overall best fit of the signal was modelled with power-laws $dN_\nu/dE_\nu=\Phi_0 E_\nu^{-\Gamma}$, yielding relatively soft spectral indices ($\Gamma>2$). The value $\Gamma=2$ is expected for neutrinos produced from primary CRs accelerated by the simplest Fermi shock acceleration models \cite{bib:fermiacc, bib:fermiacc2} and interacting near their sources \cite{bib:waxman}. The recent IceCube analysis \cite{bib:ICcombi} using different data samples excludes the $E_\nu^{-2.0}$ spectrum in the energy range between 25 TeV and 2.8 PeV with a significance of more than 3.8$\sigma$, assuming that the astrophysical neutrino flux is isotropic and consisting of equal flavours at Earth. Under the same assumptions, the best-fit spectral index is $\Gamma=2.50\pm 0.09$ and the normalization at the energy of 100 TeV (for all three neutrino flavours, $3f$) is $\Phi^{3f}_0(100\ \textrm{TeV}) = 6.7^{+1.1}_{-1.2}\cdot 10^{-18}$\,GeV$^{-1}$\,cm$^{-2}$\,s$^{-1}$\,sr$^{-1}$. No significant excess is found when searching for spatial anisotropies. 

IceCube observes neutrinos from the Southern sky as downward-going events. The HESE sample is selected with a vetoing technique that disfavours the detection of track-like events. As a consequence, the event sample is dominated by shower-like events, for which the detector has a typical angular uncertainty of 10-20$^\circ$. Muon neutrinos coming from the Northern hemisphere are detected as upward-going muon tracks from charged current (CC) interactions \cite{bib:ICmu}, yielding a best-fit, single-flavour flux $\Phi^{1f}_0(100\ \textrm{TeV}) = 9.9^{+3.9}_{-3.4}\cdot 10^{-19}$\,GeV$^{-1}$\,cm$^{-2}$\,s$^{-1}$\,sr$^{-1}$ and assuming $\Gamma=2$.

The separate fit of the fluxes from the Northern and Southern hemispheres \cite{bib:ICcombi} indicates a preference (although with small statistical significance) for a harder spectrum in the Northern hemisphere. Moreover, some authors have observed that events are concentrated near the Galactic Centre and Galactic Plane regions in a way that seems inconsistent with an isotropic neutrino distribution \cite{bib:neroEvidence,bib:vissani}. Such a difference between the Northern and Southern skies could potentially stem from the presence of a softer contribution to the neutrino flux from the Galaxy in the Southern hemisphere \cite{bib:maurizio}.

The isotropic distribution of extragalactic sources (such as active galactic nuclei or $\gamma$-ray bursts) presumably dominating the signal from the Northern hemisphere. Models generally foresee that neutrinos from these sources will be generated via photo-hadronic interactions of high-energy protons with low-energy photons of the background. These models are characterised by relatively high-energy thresholds (due to charged pion production) and disfavour a soft neutrino spectrum \cite{wax,prot}.

Neutrinos produced by proton-proton (or nuclei) interactions through the decay chain of secondary charged mesons (mainly pions) have a spectrum with a spectral index $\Gamma$ close to that of the parent hadrons but with a lower energy threshold \cite{ahro}. Since in p-p interactions the number of charged pions is approximately twice that of neutral pions (which decay to a pair of $\gamma$), the neutrino flux can be constrained from the observed $\gamma$-ray flux. Due to the high density of matter in the central part of the Galactic Plane, a neutrino signal coming from this part of the sky, mostly located in the Southern hemisphere, is expected to follow this emission scenario. 

Fermi-LAT data provide the best measurement of the diffuse $\gamma$-ray flux in the Galactic Plane up to $\sim$\,100\,GeV \cite{bib:fermigal}. Given certain model assumptions, the fraction of this flux attributed to hadronic processes can be estimated, allowing the derivation of the neutrino yield from CR propagation. Conventional models of CR propagation in our Galaxy predict a much lower and softer neutrino spectrum ($\Gamma\simeq 2.7$) \cite{joshi,kache} than that measured by IceCube. 

New predictions for the neutrino production due to CR propagation have been presented recently.
The authors of \cite{bib:antonio} start with the observation that conventional models of Galactic CR propagation (in a medium normalised to the locally observed one, and with the same diffusion properties in the whole Galaxy) cannot explain the large $\gamma$-ray flux measured by Milagro \cite{milagro} from the inner Galactic Plane region and by H.E.S.S. \cite{hess} from the Galactic Ridge region. To reconcile Fermi-LAT, Milagro and H.E.S.S. data, they have developed a phenomenological model characterised by radially-dependent CR transport properties, which foresees a neutrino spectral index in the range $\Gamma\simeq 2.4 - 2.5$.
In \cite{bib:fresh}, a sizeable neutrino flux is expected to be produced by the interaction of fresh CRs, which are hadrons supplied by young accelerators and contained by the local magnetic field, with the ambient matter.
The authors of \cite{bib:neronov} note that IceCube observes 3 events in the $E_\nu >$ 100\,TeV energy range with arrival direction compatible with a Galactic Ridge origin ($|\ell|<30^\circ$, $|b|<4^\circ$). Furthermore, the corresponding neutrino flux matches the high-energy power-law extrapolation of the spectrum of diffuse $\gamma$-ray emission from the Galactic Ridge as observed by Fermi-LAT. This motivates the hypothesis that these IceCube neutrino events and Fermi-LAT $\gamma$-ray flux are both produced in interactions of CRs with the interstellar medium in the inner Galactic region. All these models predict an enhancement of the neutrino flux coming from a limited region close to the Galactic Centre. 

\section{The ANTARES detector and dataset} \label{sec:detector}

The ANTARES underwater neutrino telescope \cite{bib:antares} is located 40 km off the Southern coast of France in the Mediterranean Sea (42$^\circ$ 48$^\prime$ N, 6$^\circ$ 10$^\prime$ E). It consists of a three-dimensional array of 10-inch photomultiplier tubes (PMTs). Neutrino detection is based on the observation of Cherenkov light induced in the medium by relativistic charged particles. Some of the emitted photons produce signals in the PMTs (``hits''). The position, time and collected charge of the hits are used to infer the direction and energy of the incident neutrino.

The study presented here focuses on track-like events, associated with CC interactions of muon neutrinos. The muon direction is correlated with that of the incoming neutrino, and a sub-degree angular resolution on the neutrino arrival direction can be achieved by means of a maximum likelihood fit \cite{anta12}. 

Data collected from May 2007 to December 2013 constitute the data sample for the present analysis, with an effective total livetime of 1622 days. High quality data runs, defined according to environmental and data taking conditions, have been selected for this work (analogously to \cite{bib:antares_ps}). A detailed Monte Carlo simulation is available for each data acquisition run \cite{bib:simu, bib:rbr}.

\section{The search method} \label{sec:data_analysis}

An enhancement of the neutrino diffuse emission from a region of the sky covering a small solid angle can be searched for by comparing the number of events coming from the region (on-zone) to that of regions with no expected signal and the same acceptance to the background (off-zones). To enhance the harder signal over the background of atmospheric neutrinos, a cut selecting mainly high-energy events is defined. This approach has already been used to search for neutrino candidates from the region of the Fermi Bubbles \cite{fb}.
Optimising this method requires: 1) an efficient suppression of atmospheric events; 2) the optimisation of the size of the search region and 3) the subsequent definition of background-only regions, each having the same exposure as that of the signal region.
The analysis uses Monte Carlo simulations only in the optimisation of the event selection; this avoids biases in the estimation of the signal and background and reduces systematic effects. Monte Carlo data sets are produced simulating real data acquisition conditions, taking into account the actual detection efficiency of the apparatus.

The signal is assumed to be a power-law diffuse flux with arbitrary normalisation and spectral indices varying from $\Gamma=2.0$ to $2.7$. 
Motivated by the IceCube best fit and models of neutrino production from CR propagation, the event selection criteria have been optimised in order to achieve the best sensitivity for a signal with spectral index $\Gamma=2.4$. They are identical to those obtained for $\Gamma=2.5$.
The optimal cuts are found using the Model Rejection Factor (MRF) minimisation technique \cite{bib:mrf}. 

The background component due to mis-reconstructed atmospheric muons, which mimick upgoing neutrino events, has been simulated using the MUPAGE program \cite{bib:mupage}. This background is suppressed by cuts on quality parameters of upgoing reconstructed tracks: $\Lambda$, which is related to the maximum likelihood of the fit, and $\beta$, which estimates the angular error. The distributions of $\Lambda$ and $\beta$ for atmospheric neutrinos, atmospheric muons and data are reported in \cite{anta12}. It is found that the cut $\Lambda > -5.0$ and $\beta < 0.5^\circ$ optimises the MRF and suppresses the contamination from wrongly reconstructed atmospheric muons in the upgoing sample to the level of 1\%.

The remaining background consists of atmospheric neutrinos \cite{an:nu_flux}. The \textit{conventional} component, coming from the decay of pions and kaons, has been shaped according to \cite{bib:honda} while the flux from \cite{bib:enberg} has been used for the \textit{prompt} component, expected from charmed hadron decays. This component is reduced by imposing a cut on the estimated energy of the events, limiting the event sample to the energy where the harder cosmic flux is expected to emerge above the atmospheric background. For this analysis the energy estimator E$_{\rm{ANN}}$ \cite{bib:jutta}, derived from an artificial neural network algorithm, is used. The standard deviation of the variable $\log_{10} (E_{\rm{ANN}}/E_{\rm{true}})$, where $E_{\rm{true}}$ is the Monte Carlo true energy of the muon, is almost constant at $\sim 0.4$ over the considered energy range. The MRF optimisation results in $ E_{\rm{ANN}}^{\rm{cut}} = 10$\,TeV as the best cut value. Above $ E_{\rm{ANN}}^{\rm{cut}}$, only 6\% of the selected atmospheric neutrinos survive while 40\% of the signal (for $\Gamma=2.4$) passes the cut.

Assuming a direct connection between the emission of $\gamma$-rays and neutrinos from  pion decay in hadronic mechanisms \cite{Visser}, the $\gamma$-ray flux measured by Fermi-LAT \cite{bib:fermigal} is used to estimate the flux of galactic neutrinos. Though this diffuse emission is extended over the whole Galactic Plane, it is much brighter in the very central region; including non-central regions of the plane in this search would mostly increase the atmospheric background. The MRF method is used to determine the optimal search region for each spectral index. For a signal spectrum with $\Gamma=-2.4$, the signal region is represented by the rectangle (enclosing the Galactic Centre) in galactic coordinates with longitude $|\ell|< 40^\circ$ and latitude $|b|< 3^\circ$. This corresponds to a solid angle of $\Delta \Omega= 0.145$\,sr. Modifications to the longitudinal size of the signal region do not significantly reduce the resulting sensitivity, while the latitude bound has a larger effect - about 10\% worsening per degree of increased size. 

\begin{figure}
  \centering
  \includegraphics[width=0.7\textwidth]{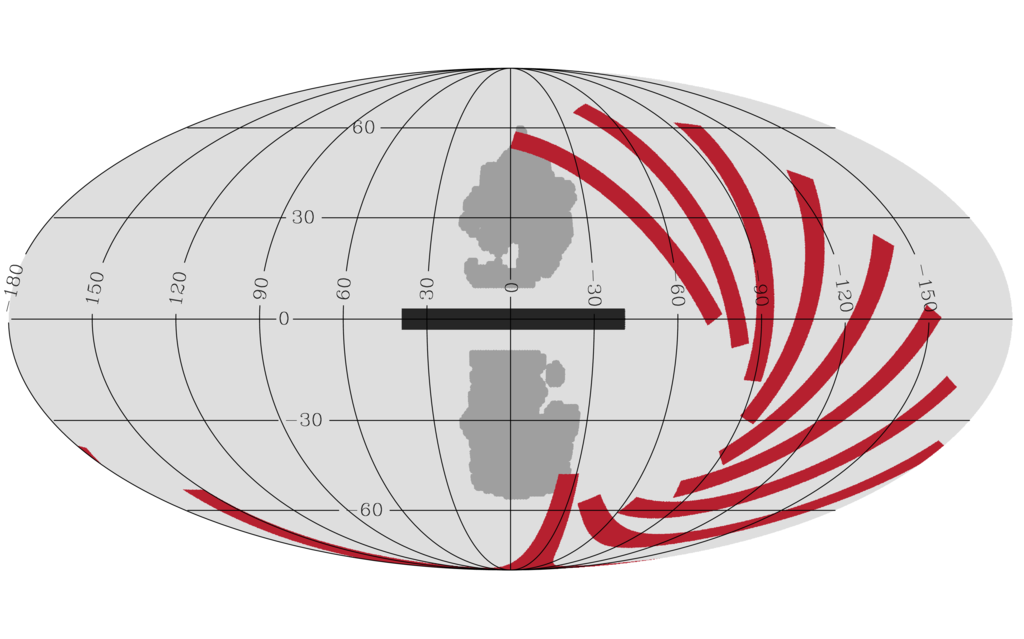}
  \caption{Aitoff projection in galactic coordinates of the signal (black) and background (red) regions, representing the considered Galactic Plane region and off-zones of the analysis. Also shown are the Fermi Bubbles (grey) as in \cite{bib:fb_map}.}
  \label{fig:onoff}
\end{figure}

Off-zones are defined as fixed regions in equatorial coordinates, which have identical size and shape as the signal region and are not overlapping with it or each other. In local coordinates, off-zones span the same fraction of the sky as the on-zone, but with some fixed delay in time, i.e. they differ only in right ascention. They are shifted in the sky to avoid any overlap with the Fermi Bubble regions \cite{bib:fb_map}, so that none of the possible signal events from these areas enters into the background estimation. The maximum number of independent off-zone regions is 9. The  signal and background regions in galactic coordinates are shown in Figure \ref{fig:onoff}. Data from the signal region were blinded until the event selection procedure was completely defined. Off-zones can also be used to test the agreement between data and Monte Carlo.

After the optimisation procedure, considering a signal flux with an energy spectrum with $\Gamma=-2.4$ ($-2.5$) the expected limit at 90\% confidence level (c.l.) for the considered data sample corresponds to $\Phi_0^{1f}$(1 GeV) = 2.0 (6.0)$\cdot$10$^{-5}$\,GeV$^{-1}$\,cm$^{-2}$\,s$^{-1}$\,sr$^{-1}$. For the normalisation at a different energy E, the fluxes must be multiplied by the factor $\left(\frac{\textrm{E}}{\textrm{1 GeV}}\right)^{-\Gamma}$. For all flavours, the normalisation must be multiplied by a factor three under the assumption of a cosmic flux in flavour equipartition ($\nu_e : \nu_\mu : \nu_\tau = 1:1:1$). 
The energy range between 3 and 300\,TeV contains the central 90\% of the expected detected signal.

\section{Results} \label{sec:results}

\begin{figure}
  \centering
  \includegraphics[width=0.7\textwidth]{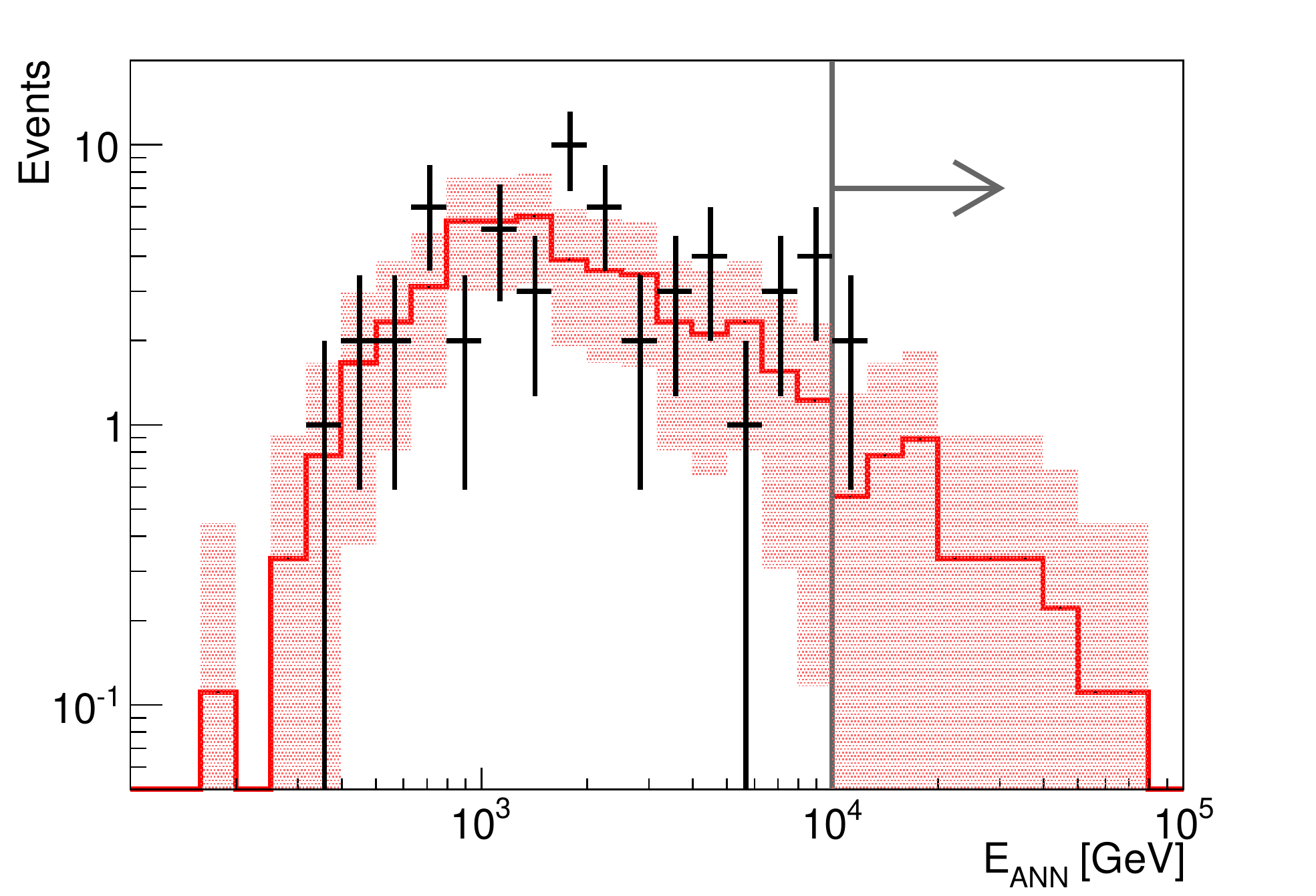}
  \caption{Distribution of the reconstructed energy $E_{\rm{ANN}}$ of upgoing muons in the Galactic Plane (black crosses) and average of the off-zone regions (red histogram). The grey line shows the energy selection cut applied in the procedure.}
  \label{fig:energy}
\end{figure}

After the unblinding of the entire data sample, 3.7 events surviving cuts are observed on average in the off-zone regions, while two are detected from the Galactic Plane region. The distributions of the number of selected events in the on-zone and off-zone regions as a function of the reconstructed energy are reported in Figure \ref{fig:energy}.

A smaller number of events is observed in the signal region than the expected background, and the Feldman and Cousins 90\% c.l.\,upper bound \cite{bib:F&C} is computed. For $\Gamma$ = 2.4 the corresponding flux $\Phi_0^{1f}$(1 GeV) = $1.5\cdot 10^{-5}$\,GeV$^{-1}$\,cm$^{-2}$\,s$^{-1}$\,sr$^{-1}$. However, adopting the same conservative approach as for the limits from selected point-like sources \cite{bib:antares_ps} in the case of an underfluctuation, the 90\% c.l.\,upper limit on the signal flux is set to the value of the ANTARES sensitivity. One limit for each considered spectral index is obtained.

\begin{figure}
  \centering
  \includegraphics[width=0.8\textwidth]{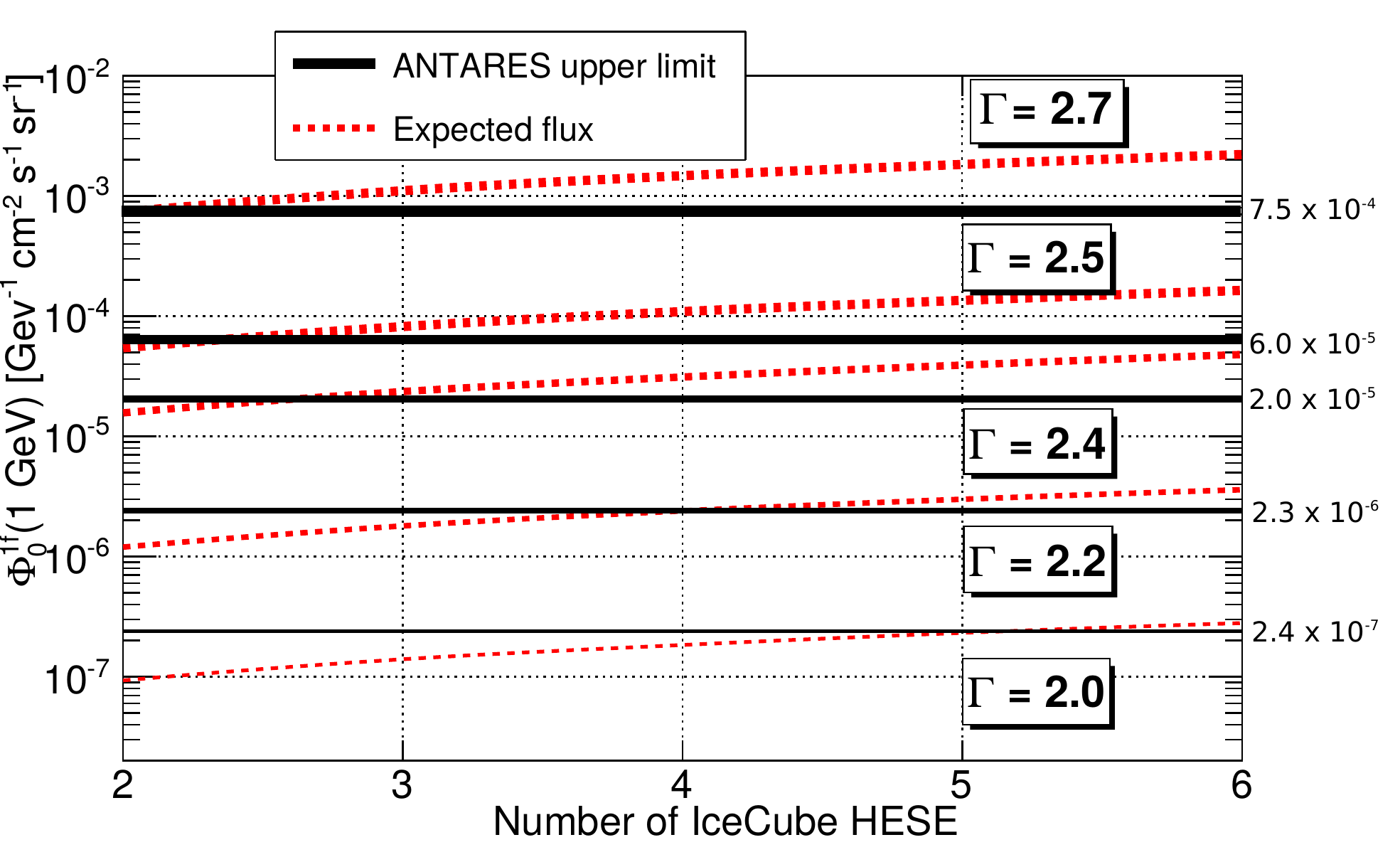}
  \caption{ANTARES upper limits (black) derived for the Galactic Plane region for different signal spectral indices $\Gamma$, compared to the flux required to produce from 2 to 6 IceCube HESE in the signal region (red dashed lines). 
Selection cuts have been optimised for $\Gamma$ = 2.4 and 2.5. The limits for softer and harder spectral indices are thus derived with non-optimal criteria. }
  \label{limits}
\end{figure}

The 90\% c.l.\,upper limits on $\Phi_0^{1f}$(1 GeV) are reported in Figure \ref{limits} for particular values of $\Gamma$. For each value of $\Gamma$, the one-flavour neutrino flux from the considered region necessary to produce from 2 to 6 HESE is also reported. The curves are computed on the basis of the effective areas reported in \cite{bib:IC2y} according to the prescription of \cite{bib:maurizio}. All fluxes above the horizontal black lines are excluded at 90\% c.l.\,by ANTARES observation. For instance, a flux with spectral index $\Gamma=2.5$ that produces 3 or more HESE in the signal region of $\Delta \Omega=0.145$\,sr is excluded. For the conventional CR propagation scenario, the 90\% c.l.\,upper limit for $\Gamma$\,=\,2.7 is $\Phi_0^{1f}$(1 GeV) = 7.5$\cdot$10$^{-4}$\,GeV$^{-1}$\,cm$^{-2}$\,s$^{-1}$\,sr$^{-1}$.

\begin{figure}
  \centering
  \includegraphics[width=0.9\textwidth]{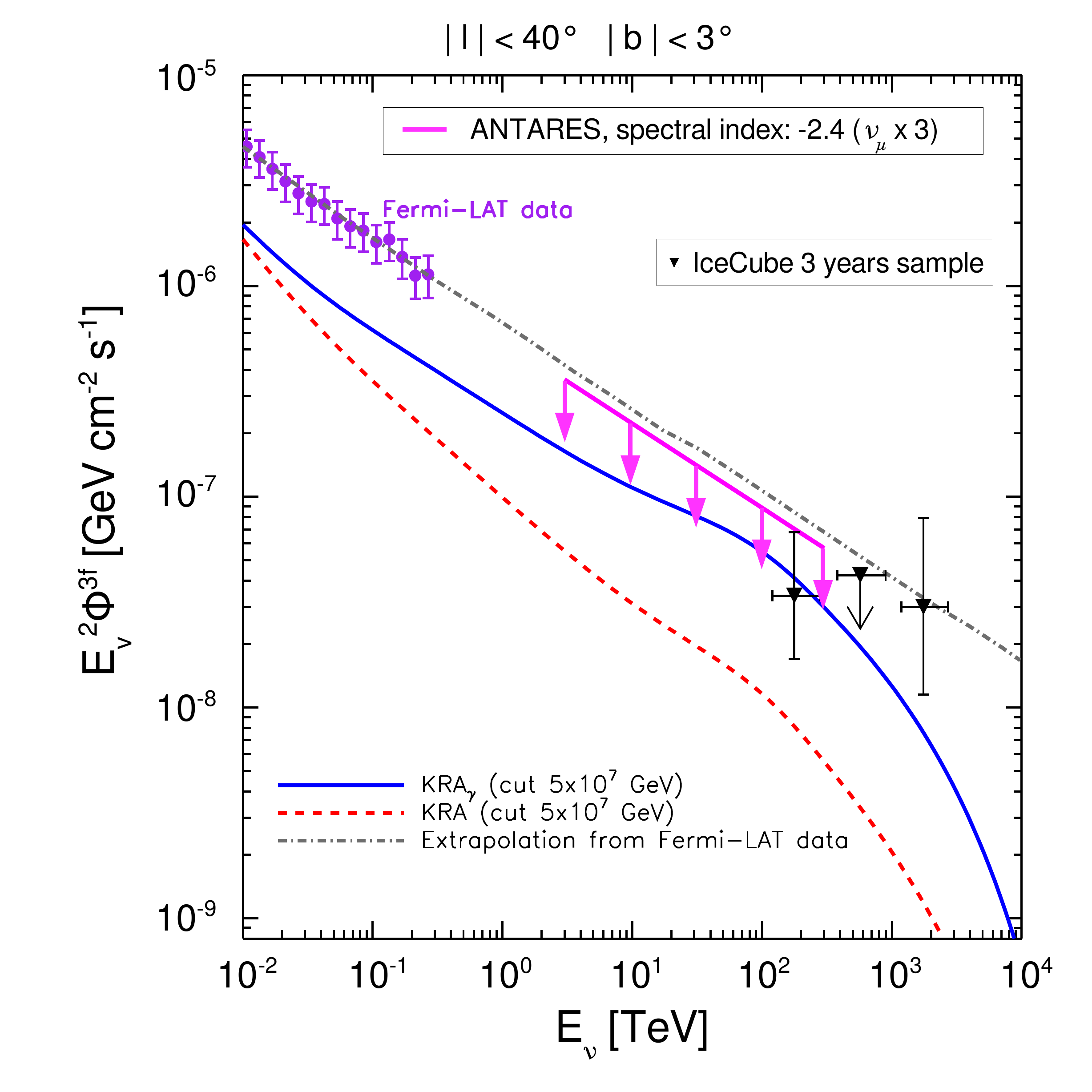}
    \caption{ANTARES upper limit (magenta line) on the neutrino flux from the Galactic Plane region ($|\ell|<40^\circ$, $|b|<3^\circ$) compared to expectations as computed in \cite{bib:antonio}, assuming a CR cut-off at 5$\times10^7$\,GeV, both with (KRA$_\gamma$) and without (KRA) spectral hardening.
The neutrino flux (dot-dashed line) extrapolated from the Fermi-LAT diffuse $\gamma$ flux adapted from \cite{bib:neronov} up to IceCube energies is shown. The implied flux from the three events from the IceCube 3 years sample \cite{bib:IC3y} is shown as black triangles.}
    \label{result}
\end{figure}

Figure \ref{result} shows the computed ANTARES 90\% c.l.\,upper limit for the neutrino emission in the region $|\ell|< 40^\circ$ and $|b|< 3^\circ$ assuming a $\Gamma=2.4$ neutrino flux. The limit on $\Phi_0^{3f}$ assuming flavour equipartition is reported, along with expectations from models. 
The simple extrapolation \cite{bib:neronov} to IceCube energies of the diffuse $\gamma$-ray flux measured by Fermi-LAT \cite{bib:fermigal} is excluded at 90\% confidence level, assuming flavour equipartition. 
Models (KRA$_\gamma$, Figure \ref{result}) that consider a harder CR spectrum in the inner Galaxy, and the hardening of the CR spectrum measured by PAMELA and AMS-02 \cite{bib:antonio}, yield a neutrino flux (at 100\,TeV) of a factor of two to three lower.
Models not including the CR hardening (KRA, Figure \ref{result}) yield neutrino fluxes one order of magnitude smaller than that of the extrapolation from Fermi-LAT.


\section{Conclusions and outlook} \label{sec:outlook}

An enhanced neutrino production from the central part of the Galactic Plane has been searched for using track-like events observed by the ANTARES telescope from 2007 to 2013. No excess of events has been observed, and limits on the contribution from this possible source to the astrophysical neutrino signal observed by IceCube have been set as a function of spectral index. For a neutrino flux $\propto$E$^{-2.5}$ we exclude at 90\% c.l.\,that 3 or more events from the 3 year IceCube HESE sample are originating from this region.
The extrapolation of the Fermi-LAT $\gamma$-ray measurement to the IceCube neutrino flux in the Galactic Plane area has also been constrained.
 
Data taking of the ANTARES neutrino telescope will continue at least up to the end of 2016, increasing the $\nu_\mu$ statistics available for this analysis. In addition, a new reconstruction procedure for showering events has been developed, with an angular resolution of 3-4 degrees in the TeV-PeV range \cite{bib:showers_icrc}, which can be used to enhance the sensitivity for point-like sources and diffuse emission from small regions of the sky. Preliminary results indicate that using reconstructed cascades, the sensitivity to point sources with $\Gamma$=2 spectrum improves by about 30\%. This suggests that at the end of data taking the sensitivity of ANTARES will reach a level close to the prediction of the model that includes a CR spectral hardening (KRA$_\gamma$) \cite{bib:antonio}.
 

\section*{Acknowledgements}

We are indebted to D. Gaggero, D. Grasso, A. Urbano and M. Valli for the useful discussion, comments and suggestions.
The authors acknowledge the financial support of the funding agencies: Centre National de la Recherche Scientifique (CNRS), Commissariat \`a
l'\'ener\-gie atomique et aux \'energies alternatives (CEA), Commission Europ\'eenne (FEDER fund and Marie Curie Program), Institut Universitaire de France (IUF), IdEx program and UnivEarthS Labex program at Sorbonne Paris Cit\'e (ANR-10-LABX-0023 and ANR-11-IDEX-0005-02), R\'egion \^Ile-de-France (DIM-ACAV), R\'egion Alsace (contrat CPER), R\'egion Provence-Alpes-C\^ote d'Azur, D\'e\-par\-tement du Var and Ville de La Seyne-sur-Mer, France; Bundesministerium f\"ur Bildung und Forschung (BMBF), Germany; Istituto Nazionale di Fisica Nucleare (INFN), Italy; Stichting voor Fundamenteel Onderzoek der Materie (FOM), Nederlandse organisatie voor Wetenschappelijk Onderzoek (NWO), the Netherlands; Council of the President of the Russian Federation for young scientists and leading scientific schools supporting grants, Russia; National Authority for Scientific Research (ANCS), Romania; Mi\-nis\-te\-rio de Econom\'{\i}a y Competitividad (MINECO), Prometeo and Grisol\'{\i}a programs of Generalitat Valenciana and MultiDark, Spain; Agence de  l'Oriental and CNRST, Morocco. We also acknowledge the technical support of Ifremer, AIM and Foselev Marine for the sea operation and the CC-IN2P3 for the computing facilities.

\end{document}